\documentclass[11pt]{article}
\usepackage{hyperref}
\pdfoutput=1

\usepackage[english]{babel}
\usepackage{amsmath}
\usepackage{amssymb}
\usepackage{graphicx}
\usepackage{natbib}
\usepackage[nolists,tablesfirst, noheads]{endfloat}
\usepackage{appendix, bbm, color}
\usepackage{fullpage, multirow}
\usepackage{authblk}
\usepackage{tikz}
\usetikzlibrary{fit,positioning,calc}

\providecommand{\keywords}[1]{\textbf{\textit{Keywords:}} #1}
\def\bmu{\mbox{\boldmath $\mu$}}
\def\bnu{\mbox{\boldmath $\nu$}}

\def\bgamma{\mbox{\boldmath $\gamma$}}

\def\bpsi{\mbox{\boldmath $\psi$}}
\def\btheta{\mbox{\boldmath $\theta$}}

\def\bLambda{\mbox{\boldmath $\Lambda$}}

\def\bbeta{\mbox{\boldmath $\beeta$}}

\def\bzeta{\mbox{\boldmath $\zeta$}}
\def\beeta{\mbox{\boldmath $\eta$}}
\def\bSigma{\mbox{\boldmath $\Sigma$}}

\def\bSigma{\mbox{\boldmath $\Sigma$}}
\def\bZero{\mbox{\boldmath $0$}}
\def\bI{\mbox{\boldmath $I$}}

\def\bmu{\mbox{\boldmath $\mu$}}

\def\bgamma{\mbox{\boldmath $\gamma$}}

\def\bDelta{\mbox{\boldmath $\Delta$}}

\def\bZ{\mathbf{Z}}

\def\bbb{\mathbf{b}}

\def\bZ{\mathbf{Z}}

\def\bx{\mathbf{x}}
\def\bX{\mathbf{X}}

\begin{document}
\title{Spatial Bayesian Latent Factor Regression Modeling of Coordinate-­based Meta­-analysis Data}
\author[1]{Silvia Montagna\thanks{S.Montagna@warwick.ac.uk}}
\author[2]{Tor Wager}
\author[3]{Lisa Feldman-Barrett}
\author[4]{Timothy D. Johnson}
\author[1]{Thomas E. Nichols}
\affil[1]{Department of Statistics, University of Warwick, CV4 7AL Coventry, UK}
\affil[2]{Department of Psychology and Neuroscience, University of Colorado at Boulder, Boulder, CO 80309, U.S.A.}
\affil[3]{Department of Psychology, Northeastern University, Boston, MA 02115, U.S.A.}
\affil[4]{Biostatistics Department, University of Michigan, Ann Arbor, MI 48109, U.S.A.}

\renewcommand\Authands{ and }

\maketitle

\begin{abstract}
Now over 20 years old, functional MRI (fMRI) has a large and growing literature that is best synthesised with meta-analytic tools. As most authors do not share image data, only the peak activation coordinates (foci) reported in the paper are available for Coordinate-based Meta-analysis (CBMA). Neuroimaging meta-analysis is used to 1) identify areas of consistent activation; and 2) build a predictive model of task type or cognitive process for new studies (reverse inference). To simultaneously address these aims, we propose a Bayesian point process hierarchical model for CBMA. We model the foci from each study as a doubly stochastic Poisson process, where the study-specific log intensity function is characterised as a linear combination of a high-dimensional basis set. A sparse representation of the intensities is guaranteed through latent factor modeling of the basis coefficients. Within our framework, it is also possible to account for the effect of study-level covariates (meta-regression), significantly expanding the capabilities of the current neuroimaging meta-analysis methods available. We apply our methodology to synthetic data and a neuroimaging meta-analysis dataset. 
\end{abstract}


\section{Introduction}
\label{s:intro}
Functional magnetic resonance imaging (fMRI) has become an essential, non-invasive, tool for learning patterns of activation in the working human brain (e.g., \cite{ Pekar2006, Wager2015}).  Whenever a brain region is engaged in a particular task, there is an increased demand for oxygen in that region which is met by a localised increase in blood flow. The MRI scanner captures such changes in local oxygenation via a mechanism called the {\textit{Blood Oxygenation Level-Dependent} (BOLD)} effect; see, e.g., \cite{Buxton2007} for a brief introduction on fMRI. The great popularity that fMRI has achieved in recent years is supported by various software packages that implement computationally efficient analysis through a mass univariate approach (MUA). Specifically, MUA consists of fitting a general linear regression model at each voxel independently of every other voxel, thus producing images of parameter estimates and test statistics. These images are then thresholded to identify significant voxels or clusters of voxels, and significance is determined via random field theory \citep{Evans1996} or permutation methods \citep{Holmes2001}. Despite its simplicity, the MUA lacks an explicit spatial model. Even though the activation of nearby voxels is correlated estimation with the MUA ignores the spatial correlation, but inference later accounts for it when random field theory or permutation procedures define a threshold for significant activation. \\
\indent The relatively high cost of MRI scanner time, however, pose some limitations to single fMRI studies. The main limitation is the small number of subjects that can be recruited for the study, often fewer than 20 \citep{Carp2012}. As a result, most fMRI studies suffer from inflated type II errors (i.e., low power) and poor reproducibility \citep{Thirion2007}. To overcome these limitations there has been an increasing interest in the meta-analysis (MA) of neuroimaging studies. By combining the results of independently conducted studies, MA increases power and can be used to identify areas of consistent activation while discounting chance findings. \\ 
\indent In addition to the identification of areas of consistent activation (forward inference, or Pr[Activation $\vert$ Task]), there has been intense interest in the development of meta-analytic methods to implement proper reverse inference \citep{Yarkoni2011}. Reverse inference refers to inferring which cognitive process or task generated an observed activation in a certain brain region (Pr[Task $\vert$ Activation]). Suppose that researchers develop a task to probe cognitive process $A$ and find that brain area $X$ is activated. A common but misguided practice in neuroimaging is to conclude that activation of brain region $X$ is evidence that cognitive process $A$ is engaged. However, this logic is wrong and the resulting inference is faulty. In fact, it is known that a single region may be activated by several different tasks \citep{Yeo2015}. \\
\indent Given that published fMRI studies rarely share the statistic images or raw data, MA techniques are typically based on coordinates of activation, that is, the $(x, y, z)$ coordinates of local maxima in significant regions of activation, where the coordinate space is defined by a standard anatomical atlas. We shall refer to these coordinates as foci (singular focus), and denote the MA based on foci as Coordinate-Based Meta-Analysis (CBMA). Several approaches to CBMA can be found in the literature. See, for example, \cite{Fox1997, Nielsen2002, Turkeltaub2002, Wager2004, Kober2008a, Eickhoff2009, Radua2009, Kang2011, Yue2012, Kang2014}. These methods can be categorised as either kernel-based or model-based approaches.\\  
\indent The most popular kernel-based approaches to CBMA are activation likelihood estimation (\cite{Turkeltaub2002}, ALE), modified ALE (\cite{Eickhoff2009}, modALE), and multilevel kernel density analysis (\cite{Wager2007, Kober2008a}, MKDA). These methods proceed in three main steps. First, one creates {\em focus maps} for each focus in each study; in these images the intensity at each voxel depends on the proximity of that voxel to that map's focus. For each study, there are as many focus maps as the number of reported foci. These focus maps are then combined to create study maps, which are further combined into a single statistic image (MA map) that represents the evidence for consistent activation (clustering). Significance of the statistic image is assessed with a Monte Carlo test under the null hypothesis of complete spatial randomness. The difference across the aforementioned methods lie in how they create the foci maps, and in how these maps are combined into study and MA maps. We defer to \cite{Pantelis2016} for an extensive review of these methods. These approaches, however, have some serious limitations. In particular, they are based on a MUA, and require ad-hoc spatial kernel parameters that need to be pre-specified. \\
\indent Recently, model-based approaches have been proposed to overcome some of the limitations of kernel-based methods. See \cite{Pantelis2016} for an extensive review of model-based methods. All of these methods are grounded in the spatial statistics literature and utilize spatial stochastic models for the analysis of the foci. However, there are relatively few works that take this approach. \cite{Kang2011} propose a Bayesian spatial hierarchical model using a marked independent cluster process. Despite its flexibility, the model involves many hyperprior distributions whose parameters are challenging to specify and require expert opinion, and the posterior intensity function is somewhat sensitive to the choice of hyperpriors. \cite{Yue2012} propose a Bayesian spatial binary regression model where the probability that a voxel is reported as a focus, $p(\bnu)$, is modeled via logistic regression, $p(\bnu) = \Phi(z(\bnu))$, and $z(\bnu)$ is modeled as a spatially adaptive Gaussian random field. This method, however, does not treat the number and the location of the foci for each study as random. Also, it does not treat the MA studies as the units of observation, but rather the data at each voxel. Further, both \cite{Kang2011} and \cite{Yue2012} propose models for a single, homogeneous group of studies whereas it is common practice in MA to simultaneously consider several types of tasks. To address this limitation, \cite{Kang2014} propose a Bayesian hierarchical Poisson/gamma random field model for multi-type MA by generalizing the Poisson/gamma random field model developed by \cite{Wolpert1998}. The foci from each type of study are modeled as a cluster process driven by a random intensity function that is modeled as a kernel convolution of a gamma random field. The type-specific gamma random fields are related and modeled as a realization of a common gamma random field shared by all types, inducing correlation between study types. Also, the authors propose a model-based classifier to perform reverse inference. \\
\indent In this paper, we propose a Bayesian hierarchical model that is both simpler and more flexible than previous Bayesian point process models for CBMA \citep{Kang2011, Kang2014}. In particular, we extend the Bayesian latent factor regression model for longitudinal data of \cite{Montagna2012} to the analysis of CBMA data. We model the foci from each study as a ``doubly stochastic" Poisson process (Cox process) \citep{Moller2004}, where the study-specific log intensity function is characterised as a linear combination of a 3­-dimensional basis set. We adaptively drop unnecessary bases by imposing sparsity on their coefficients through a latent factor regression model, and information on covariates is incorporated through a simple linear regression model on the latent factors. By interpreting the latent factors as vectors of coefficients, our construction becomes analogous to a functional principal component analysis representation of the log intensities, but in our construction bases are no longer mutually orthogonal functions. The number of latent bases is estimated via posterior computation. The latent factorisation is used as a vehicle to link the intensity functions to a study-­type as part of a scalar-­on-­image regression. Specifically, suppose the MA dataset can be split into studies of two types, A and B. We build a probit regression model to predict the study type, where the probability that study $i$ is of type A is expressed as a function of the latent factors. Along the same lines, we can easily accommodate prediction of two or more types of studies. Our fully Bayesian CBMA model permits explicit calculation of a posterior predictive distribution for study type and, as a result, allows inference on the most likely domain for any new experiment by just using its foci. \\
\indent To demonstrate the application of our algorithm to real-world data, we compared studies of emotion and cognitive control, using hand-coded activation coordinates from previous meta-analyses \citep{Kim2012, Kober2008a, Lindquist2012, Nee2007, Jonides2013, Rottschy2012, Smith2003, Wager2004}. Each domain has been studied extensively using neuroimaging in hundreds of published studies. There is substantial convergence about the systems broadly involved in each, and though they interact, cognitive control and emotion are associated with distinct large-scale networks. In addition, there is substantial converging evidence on the cognitive and emotional functions of homologous systems in invasive animal studies, further validating the functional roles of the systems identified in human neuroimaging. We aimed to both characterize the patterns of brain activation that are typical to each type, as well as evaluate the ability of our model to conduct a limited ``reverse inference'', that is, to predict which topic (cognitive control or emotion) a new study investigates. \\
\indent The remainder of this paper is organised as follows. Section~\ref{s:outline} describes our spatial latent factor regression model for CBMA data and outlines a connection with fPCA. In Section~\ref{neuroapp}, we apply our model to the MA dataset of emotion and executive control studies, and compare our results with MKDA. Section~\ref{s:simulation} reports results from sensitivity analyses and simulation studies. We conclude the manuscript with a final discussion of our model (Section~\ref{s:discuss}). An outline of the MCMC algorithm for posterior computation and an additional simulation study are reported in the Supplementary Materials.

\section{Spatial Bayesian latent factor regression for CBMA}
\label{s:outline}
In this Section, we present our spatial Bayesian latent factor regression for CBMA data. Articles often report results from different statistical comparisons called contrasts, hereafter called studies. Following the convention of existing neuroimaging CBMA, we treat the studies as independent. The model outlined in Section \ref{s:model} generalizes \cite{Montagna2012} to the case where observations are spatial point patterns from different studies. Each spatial point pattern is assumed to be an independent realization of a spatial point process. In Section \ref{s:inverseinference} we show how the model accommodates reverse inference. Section \ref{s:fPCA} further discusses the methodogoloy by presenting an analogy with functional principal component analysis (fPCA). Section \ref{s:priorelicitation} presents the prior specification for the model parameters.

\subsection{The model}
\label{s:model}
Consider independent spatial point patterns arising from $n$ studies, $\bx_1, \dots, \bx_n$. We regard $\bx_i$ as a realization of a doubly stochastic Poisson process \citep{Cox1955} $\bX_i$ driven by a non-negative random intensity function $\mu_i$ defined on a common brain template $\mathcal{B} \subset \mathbbm{R}^3$ with finite volume $\vert \mathcal{B} \vert$. Given that observations are independent, the sampling distribution is

	\begin{eqnarray}
	\nonumber \label{likelihood}
	\pi(\{\bx_i\}_{i = 1}^n \vert \{\mu_i\}_{i = 1}^n) &\propto& \prod_{i = 1}^n \left[\exp\left\{ - M_i(\mathcal{B}) \right\} \prod_{\bx_{ij	} \in \bx_i} \mu_i(\bx_{ij}) \right] \\ 
	&= & \exp\left \{ - \sum_{i = 1}^n M_i(\mathcal{B}) \right \} \prod_{i = 1}^n \prod_{\bx_{ij} \in \bx_i} \mu_i(\bx_{ij}), 
	\end{eqnarray}
	
\noindent where $\bx_i = \{ \bx_{ij}\}_{j = 1}^{n_i}$ is the set of foci reported by study $i$, $\bx_{ij} = (x_{ij1}, x_{ij2}, x_{ij3})^\top$ represents the centre of a voxel (or vertex), 
and $M_i(B)$ denotes a non-negative intensity measure, $M_i(B) = \int_B \mu_i(s) ds < \infty$, for any Borel measurable subset $B \subseteq \mathcal{B}$. To simplify the notation, we will denote a focus in the brain as $\bnu$ hereafter.\\
\indent For the modeling of the random functions $\mu_1, \dots, \mu_n$, we follow \cite{Montagna2012}. Specifically, we consider a functional representation for the (log) intensity function and write $\log \mu_i$ in terms of a collection of basis functions  
	\begin{equation}\label{logintensity}
	\log \mu_i(\bnu) = \sum_{m = 1}^p \theta_{im} b_m(\bnu) = \bbb(\bnu)^{\top}\btheta_i.
	\end{equation}
This specification implies that the $\log$ intensity function belongs to the span of a set of basis functions, $\{b_m(\cdot)\}_{m = 1}^p$, with $\btheta_i$ denoting a vector of study-specific coefficients. Choosing the functions $\{b_m(\cdot)\}_{m = 1}^p$ is particularly challenging since the appropriate basis is not known in advance and, conceptually, any bases can be chosen. For example, the B-spline basis or Gaussian kernels can be used to model smooth $\mu_i$ intensities. Hereafter, we use 3D isotropic Gaussian kernels
	\begin{equation}\label{Gaussianbasis}
	b_1(\bnu)  = 1, \quad \text{and} \quad b_{m}(\bnu) = \exp\{-b \vert \vert \bnu - \bpsi_m \vert \vert ^2\}, \quad m = 2, \dots, p,
	\end{equation}
with kernel locations $\{\bpsi_m\}_{m=2}^p$ and bandwidth $b$ to be specified according to prior knowledge. More flexible approaches allow the number and locations of the kernels to be unknown and estimated by the sampler, at the expense of a great increase in computational cost. Hereafter, we prefer adopting a computational-savvy approach by fixing the bases, and use sensitivity analysis to help us determine reasonable choices for $p$ and kernel locations. \\
\indent The functional representation considered in \eqref{logintensity} constitutes an alternative to the typical log Gaussian Cox process prior on $\mu_i$ (LGCP, \cite{Moller1998}), which is a widely popular prior within the spatial statistics literature. As its name suggests, the LGCP is a Cox process with $\mu_i(\bnu) = \exp\{Z(\bnu)\}$, where $Z$ is modeled as a Gaussian process. The most attractive feature of this model is that it provides a flexible and relatively tractable construction for describing spatial phenomena. Inference for LGCPs is, however, a computationally challenging problem, and the main barrier is the computation of the covariance matrix of $Z$. In a typical neuroimaging application, this matrix is very large as its dimensions correspond to the number of voxels in the brain mask (typically, more than 150,000 voxels on a $2 \times 2 \times 2$ mask). Fortunately, for covariance functions defined on regular spatial grids there exist fast methods for computing the covariance based on the discrete Fourier transform \citep{Wood1994, Rue2005, Taylor2014}. A basis function representation as in \eqref{logintensity} completely removes the need of computing the covariance matrix (and its inverse), hence has a natural computational advantage over LGCPs in this regard. \\
\indent By characterising the study-specific $\log$ intensity functions by a vector of coefficients with respect to a common functional basis representation, all variation between the study-specific intensities are reflected through the variation in the vectors $\btheta_1, \dots, \btheta_n$. \cite{Montagna2012} remark that a major drawback of the basis function approach is, however, the failure to obtain a low dimensional representation of the individual intensities. Low dimensional representations are crucial when building a hierarchical model where the foci are to be linked, as predictors or outcomes, with other variables under study. In our construction, the $\mu_i$'s are represented by the long vector of coefficients $(\theta_{i1}, \dots, \theta_{ip})$. Unless the $\mu_i$'s are sparse in the chosen basis, these vectors are dense, meaning that any projection of these vectors onto a lower dimensional space results in a substantial loss of information. To obtain a low-dimensional representation of $\log \mu_i$, we follow the lead in \cite{Montagna2012} and place a sparse latent factor model \citep{Arminger1998} on the basis coefficients
		\begin{equation}
		\btheta_i = \bLambda \bbeta_i + \bzeta_i, \quad \text{with} \quad \bzeta_i \sim N_p(0, \bSigma)
		\end{equation}
where $\btheta_i = [\theta_{i1}, \dots, \theta_{ip}]^\top$, $\bLambda$ is a $p \times k$ factor loading matrix with $k\ll p$, $\bbeta_i = (\eta_{i1}, \dots, \eta_{1k})^\top$ is a vector of latent factors for study $i$, and $\bzeta_i = (\zeta_{i1}, \dots, \zeta_{ip})^\top$ is a residual vector that is independent with the other variables in the model and is normally distributed with mean zero and diagonal covariance matrix $\bSigma = \text{diag}(\sigma_1^2, \dots, \sigma_p^2)$. Vectors $\bbeta_1, \dots, \bbeta_n$ can be put in any flexible joint model with other variables of interest. For example, information from covariates $\bZ_i$ can be incorporated through a simple linear model 
	\begin{equation}
	\bbeta_i = \boldsymbol{\beta}^\top \bZ_i + \bDelta_i, \quad \text{with} \quad \bDelta_i \sim N_k(0, \bI),
	\end{equation}
where $\boldsymbol{\beta}$ is a $r \times k$ matrix of unknown coefficients, and $r$ denotes the dimension of $\bZ_i$. \\
\indent Despite the simplicity of this hierarchical linear model, the resulting structure on $\log \mu_i(\bnu)$ allows a very flexible accommodation of covariate information. Specifically, if we marginalise out $\{\btheta_i, \bbeta_i\}$, our model results in a (finite rank) GP for $\log \mu_i$ with covariate dependent mean function 
$$\mathbbm{E}[\log \mu_i(\bnu)] = \sum_{l = 1}^k \boldsymbol{\beta}_l^\top \bZ_i \phi_l(\bnu)$$ 
and common covariance function 
$$\mathbbm{C}\text{ov}\{\log \mu_i(\bnu), \log \mu_i(\bnu^\prime) \} = \sum_{l = 1}^k \phi_l(\bnu)\phi_l(\bnu^\prime) + \sum_{m = 1}^p \sigma_m^2 b_m(\bnu) b_m(\bnu^\prime),$$ 
where $\phi_l(\bnu) = \sum_{m = 1}^p \lambda_{lm}b_m(\bnu)$, and $\boldsymbol{\beta}_l$ is the $l$th column of $\boldsymbol{\beta}$. 

\subsection{Reverse Inference}
\label{s:inverseinference}
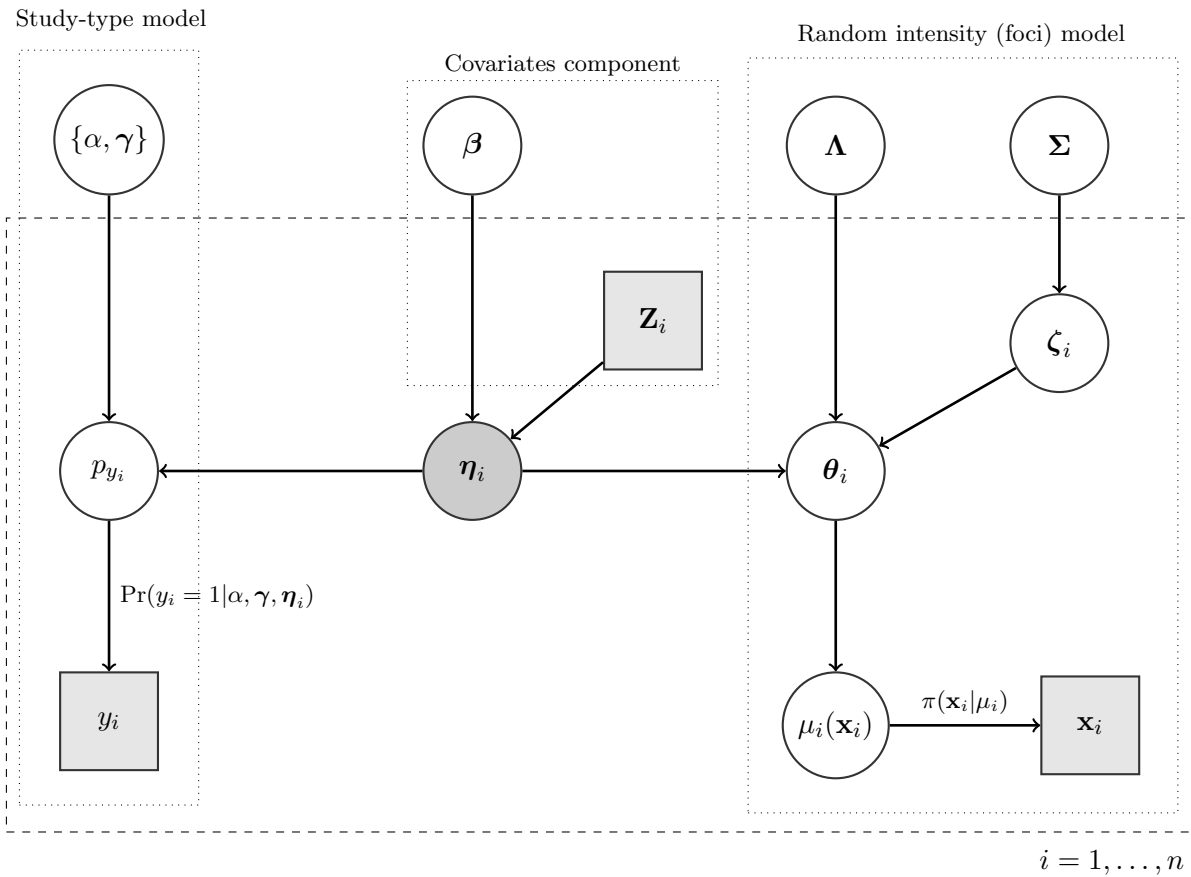
\begin{figure}[t!]
\centering
\begin{tikzpicture}
\tikzstyle{main}=[circle, minimum size = 13mm, thick, draw =black!80, node distance = 16mm]
\tikzstyle{connect}=[-latex, thick]
\tikzstyle{box}=[rectangle, draw=black!100]
\node[main, circle, fill = black!20, text centered] (eta) {$\beeta_i$};
\node[main, circle, left = 3.5 of eta, text centered] (y) {$p_{y_i}$};
\node[main, rectangle, fill = black!10, below =  2 of y, text centered] (outcome) {${y_i}$};
\node[main, right = 3.5 of eta, text centered] (theta) {$\btheta_i$};
\node[main, circle, below =  2 of theta, text centered] (px) {$\mu_i({\bx_i})$};
\node[main, rectangle, fill = black!10, right =  2 of px, text centered] (x) {$\bx_i$};
\node[main, draw, rectangle, fill = black!10, text centered] at (2.4, 2)  (Z) {$\bZ_i$};
\node[main, draw, circle, above =  3 of eta, text centered] (beta) {$\boldsymbol{\beta}$};
\node[main, draw, circle, above =  3 of theta, text centered] (Lambda) {$\bLambda$};
\node[main, draw, circle, text centered] at (7.8, 1.7) (xi) {$\bzeta_i$};
\node[main, draw, circle, above =  1.3 of xi, text centered] (Sigma) {$\bSigma$};
\node[main, draw, circle, above =  3 of y, text centered] (gamma) {$\{\alpha, \bgamma\}$};
\node[rectangle, dashed, inner sep = 7mm, draw=black!100, fit = (theta) (eta) (xi) (y) (outcome) (Z) (px) (x)] {};
\node[rectangle, dotted, inner sep = 4.5mm, draw=black!100, fit = (gamma) (y) (outcome)] {};
\node[rectangle, dotted, inner sep = 5mm, draw=black!100, fit = (theta) (Lambda) (Sigma)(x)] {};
\node[rectangle, dotted, inner sep = 2mm, draw=black!100, fit = (beta) (Z)] {};
\node[] at (8.5,-5.2) {$i = 1, \dots, n$};
\node[font=\footnotesize] at (-4.8,6) {Study-type model};
\node[font=\footnotesize] at (6.5, 5.8) {Random intensity (foci) model};
\node[font=\footnotesize] at (1.2, 5.4) {Covariates component};

\draw[->, line width = 1] (Z) --  (eta);
\draw[->, line width = 1] (beta) --  (eta);
\draw[->, line width = 1] (eta) --  (theta);
\draw[->, line width = 1] (Lambda) --  (theta);
\draw[->, line width = 1] (xi) --  (theta);
\draw[->, line width = 1] (Sigma) --  (xi);
\draw[->, line width = 1] (gamma) -- (y);
\draw[->, line width = 1] (eta) --  (y);
\draw[->, line width = 1] (y) -- node[right,font=\footnotesize]{Pr$(y_i = 1\vert \alpha, \bgamma, \bbeta_i)$}  (outcome);
\draw[->, line width = 1] (theta) --  (px);
\draw[->, line width = 1] (px) -- node[above, font=\footnotesize]{$\pi(\bx_i \vert \mu_i)$}  (x);
\end{tikzpicture}
\caption{Graphical representation of the probabilistic mechanism generating data $\{\bx_i, y_i\}$, $i = 1, \dots, n$, under the spatial Bayesian latent factor model. Shaded squares represent observed quantities and circles represent unknowns. The circle denoting the vector of latent factors is darkened. Note that the study type, $y_i$, is conditionally independent of all other nodes in the ``random intensity model'' given the latent variables $\beeta_i$. } \label{model}
\end{figure}

In response to an increasing interest in reverse inference, we focus on the development of a methodology which accommodates joint modeling of neuroimaging point pattern data with study types. Suppose we have new point pattern data $\bx_{new}$ that is a realization from one of $T$ tasks or cognitive processes, $y_{new}$. Further, we have point pattern data from $n$ studies for which the corresponding task or cognitive process is known, $\{\bx_i \vert y_i\}_{i = 1}^n$ with $y_i \in \{1, \dots, T\}$. Interest is in quantifying the probability that the new point pattern data arose from a specific task type, that is, the posterior predictive probability that $\bx_{new}$ originates from type $t$, Pr$(y_{new} = t \vert \bx_{new}, \{\bx_i \vert y_i\}_{i = 1}^n)$. Our fully Bayesian model for neuroimaging point pattern data allows inference on the most likely domain for any new experiment.\\
\indent  Hereafter, we extend the model with focus on our motivating application. The MA dataset consists of studies that can be categorized as either ``emotion'' or ``executive control'' (see Section \ref{neuroapp} for details). Therefore, the interest is in estimating the probability that newly observed point pattern data arose from either an emotion or executive control experiment. Let $y_i$ denote the study type, with 
 \[ y_i = \left\{ 
  \begin{array}{l l}
    1 & \quad \text{if study $i$ is an emotion study}\\
    0 & \quad \text{if study $i$ is an executive control study}.
  \end{array} \right.\]
Because the study type can be represented as a binary response, we can build a probit model for study type and predict the posterior probability that a new point pattern data arose from either cognitive process. Specifically, we model $p_{y_i} = $ Pr$(y_i = 1 \vert \alpha, \bgamma, \bbeta_i) = \Phi(\alpha + \bgamma^\top \bbeta_i)$, where $\Phi(\cdot)$ denotes the standard normal distribution function. Parameter $\alpha$ can be interpreted as the baseline probability that study $i$ is of the emotion type, and $\bgamma^{\top}\bbeta_i$ accounts for study-specific random deviations. Notice that the latent factors $\bbeta_i$ (Section~\ref{s:model}) are used as a vehicle to link the random intensities (thus, the foci) to the study-type. \\
\indent The intercept $\alpha$ is given a N$(m_\alpha, v_{\alpha})$ prior, with $m_\alpha = \Phi^{-1}(0.50)$ assuming emotion and executive control studies are equally likely a priori, and $\bgamma$ is a vector of unknown regression coefficients with conjugate prior distribution $\bgamma \sim \text{N}_k(\bmu_\gamma, \bSigma_\gamma)$. The full conditional posterior distributions needed for Gibbs sampling are not automatically available, but we can rely on the data augmentation algorithm of \cite{Albert1993} to facilitate the computation. We introduce independent unobservable latent variables $W_1, \dots, W_n$ and define
	\begin{equation}
	y_i = \mathbbm{1}(W_i > 0), \qquad \text{with} \qquad W_i \sim \text{N}(\alpha + \bgamma^\top \bbeta_i, 1),
	\end{equation}
so that Pr$(y_i = 1\vert \alpha, \bgamma, \bbeta_i) = \Phi(\alpha + \bgamma^\top \bbeta_i)$ by marginalizing out the latent variable $W_i$. \\ 
\indent Although the presented is for 2-class CBMA data, the proposed framework can be easily modified for joint modeling of data of many different types, e. g., the probit model for a binary outcome can be replaced by an appropriate predictive model for categorical, nominal, or continuous study features. The key idea is to use the low dimensional vectors $\bbeta_1, \dots, \bbeta_n$ in all subsequent parts of the model where one seeks to link intensities $\log \mu_1, \dots, \log \mu_n$ with other variables of interest.\\
\indent We close this Section by providing a graphical representation (Figure~\ref{model}) of the spatial Bayesian latent factor model outlined above and in \S~\ref{s:model}. The vector of latent factors plays the key role in linking the two component models for study-type and random intensities, and the study-type $y_i$ is conditionally independent of all nodes in the random intensity model given the latent factors. All parameters located outside of the dashed rectangle ($\alpha, \bgamma, \boldsymbol{\beta}, \bLambda, \bSigma$) are shared and estimated by pulling information across all studies, thus allowing for borrowing of information. If covariate information is available, covariates impact on the $\bbeta_i$'s via a linear regression model. In our motivating application, however, no covariates are available and the covariates component box is omitted by simply considering a standard multivariate normal prior on $\bbeta_i$. 

\subsection{fPCA-analogue construction}
\label{s:fPCA}
The vector of latent factors $\bbeta_i$ can also be interpreted as a coefficient vector by writing
	\begin{align}\label{logdictionary}
	\log \mu_i(\bnu) = \sum_{l = 1}^k \eta_{il} \phi_l(\bnu) + r_i(\bnu), 
	\end{align}
with 
  \begin{equation}
  \phi_l(\bnu) = \sum_{m = 1}^p \lambda_{lm}b_m(\bnu) \quad \text{and} \quad r_i(\bnu) = \sum_{m = 1}^p \zeta_{im}b_m(\bnu),
  \end{equation}
where $\{\phi_l\}_{l = 1}^k$ forms an unknown non-local basis set to be learnt from the data and $r_i$ is a function-valued random intercept. \\
\indent We recall that the GP model can be viewed as an infinite dimensional basis-function expansion. For example, the Karhunen-Lo\'eve expansion of a GP $f$ (with known covariance parameters) at $\bnu$ can be written as $f(\bnu) = \sum_{k = 1}^\infty w_k e_k(\bnu)$, where the basis functions $e_k$ are orthogonal and the coefficients $\{w_k\}$ are independent, zero-mean normal random variables. The variance of $w_k$ is equal to the $k$th largest eigenvalue. The empirical version (i.e., with the coefficients computed from a sample) is known as (functional) principal component analysis. Decomposition (\ref{logdictionary}), without $r_i(\bnu)$, is analogous to a truncated fPCA representation of $\log \mu_i(\bnu)$, however bases $\phi_1(\cdot), \dots, \phi_k(\cdot)$ are no longer mutually orthogonal within our construction. Although orthogonality enhances interpretability of the elements of the decomposition, this is not a primary concern in our application because we view the latent factorisation only as a vehicle to link the intensities (and, ultimately, the observed foci) with other variables. To highlight this difference with fPCA, we refer to $\{\phi_l\}_{l = 1}^k$ as a dictionary.\\
\indent The size $k$ and the elements of the dictionary depend on how $\bLambda$ is modeled. This discussion is deferred to the next Section.

\subsection{Prior elicitation } 
\label{s:priorelicitation}
Careful modeling of $\bLambda$ is crucial in that the factor loading matrix ultimately controls sparsity and basis selection. Following \cite{Montagna2012}, we adopt a multiplicative gamma process shrinkage (MGPS) prior on the loadings \citep{Bhattacharya2011}:
	\begin{eqnarray}\label{mgpsp} 
	\lambda_{jh} \vert \iota_{jh}, \tau_h &\sim& N(0, \iota_{jh}^{-1}\tau_h^{-1}),  \quad \iota_{jh} \sim \text{Gamma}\left(\frac{\rho}{2},\frac{\rho}{2}\right),  \quad \tau_h = \prod_{l = 1}^h \delta_l \\ 
	\delta_1 &\sim & \text{Gamma}(a_1, 1), \quad \delta_l \sim \text{Gamma}(a_2, 1), \quad l \geq 2
	\end{eqnarray}
with $j = 1, \dots, p$ and $h = 1, \dots, k$. Elements $\{\delta_l\}_{l \geq 1}$ are independent, $\tau_h$ is a global shrinkage parameter for the $h$th column of $\bLambda$ and the $\iota_{jh}$'s are local shrinkage parameters for the elements in the $h$th column. Under the choice $a_2 > 1$, the $\tau_h$'s are stochastically increasing favouring more shrinkage as the column index increases and preventing the factor splitting problem \citep{Bhattacharya2011}. The local shrinkage parameters prevent over-shrinking the nonzero loadings. Therefore, the MGPS prior shrinks a subset of the loadings strongly towards zero while retaining a sparse signal. For more details on the properties of this prior, we defer to \cite{Bhattacharya2011}. \\ 
\indent To obviate the need for pre-specifying the number of factors, we follow \cite{Bhattacharya2011} and implement an adaptive algorithm for choosing $k$. The idea behind this sampler is to strike a balance between missing important factors by choosing $k$ too small and wasting computation on an overly conservative value. At iteration $t$, one monitors the number of columns of $\bLambda$ having all elements within some pre-specified small neighbourhood of zero.  If the number of such columns drops to zero, then a column is added to $\bLambda$ by sampling the new loadings from the prior distribution, and otherwise discard the redundant columns in that the contribution of the factors is negligible. The other parameters are also modified accordingly. By proposing an oversized set of basis functions \eqref{logintensity}, we can ultimately control the complexity of our model by including or excluding a particular basis \eqref{logdictionary} based on its contribution to the likelihood of the observed data. To guarantee convergence of the chain, the adaptations are designed to satisfy the diminishing adaptation condition in Theorem 5 of \cite{Roberts2007}. \\
\indent The Bayesian specification of our LFRM for spatial point pattern data is completed by placing a Gamma prior on the residual precisions, $\sigma_m^{-2}\sim \text{Gamma}(a_\sigma, b_\sigma)$, for $m = 1, \dots, p$, and, if covariate information is available, a Cauchy prior on the matrix of coefficients $\boldsymbol{\beta}$ as follows
	\begin{equation}
	\boldsymbol{\beta}_l \sim \text{N}(0, \text{Diag}(w_{lj}^{-1})), \quad w_{lj}\sim\text{Gamma}\left(\frac{1}{2}, \frac{1}{2}\right)
	\end{equation} 
for $l = 1, \dots k$ and $j = 1, \dots, r$. A multivariate Normal prior is also a suitable choice. \\
\indent The posterior computation for our spatial LFRM proceeds via a hybrid Gibbs sampler with a Hamiltonian Monte Carlo (HMC) step \citep{Neal2011b} to update the basis function coefficients $\btheta_i$. Details are provided in the Supplementary Materials (Web Appendix A).

\section{Neuroimaging MA application}
\label{neuroapp}
\begin{figure}[h!]
\centering
\includegraphics[scale = .65]{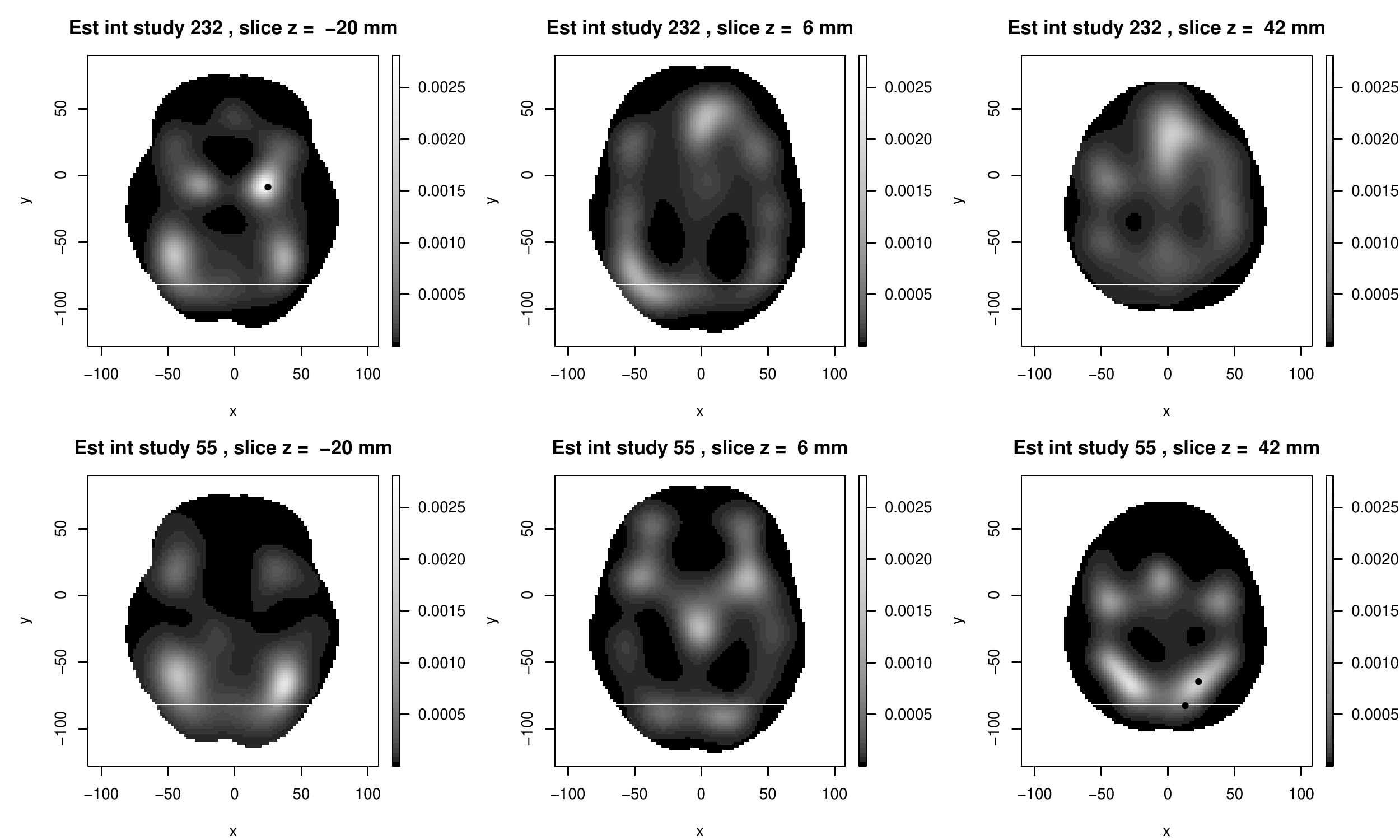}
\caption{Posterior mean intensity estimates for two randomly chosen studies. Top row: an emotion study; bottom row: an executive control study. Here we only show three axial slices (columns) of the fully 3D results. Black points denote reported foci at the corresponding axial slices. }
\label{intensities}
\end{figure}

\begin{figure}[h!]
\centering
\includegraphics[scale = .55]{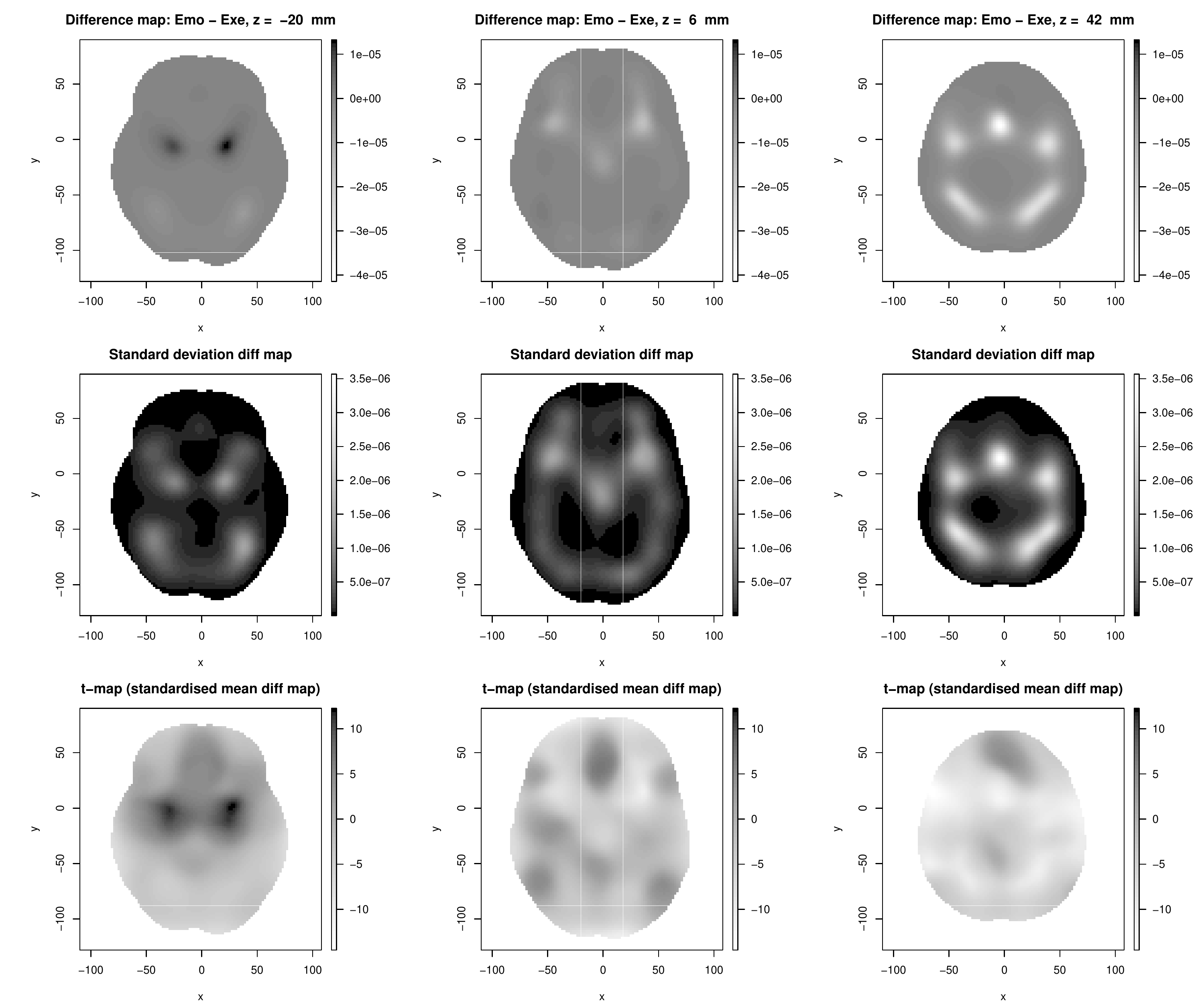}
\caption{Top row: posterior mean of difference map for the emotion studies posterior group intensity vs. executive control studies posterior group intensity; middle row: estimated standard deviation of difference map; bottom row: standardised mean difference map. Here we only show three axial slices (columns) of the fully 3D results. The grey color scale for the middle row has been reversed to improve visibility of the results. }
\label{means}
\end{figure}

\begin{figure}[h!]
\centering
\includegraphics[width = 16cm, height = 20cm]{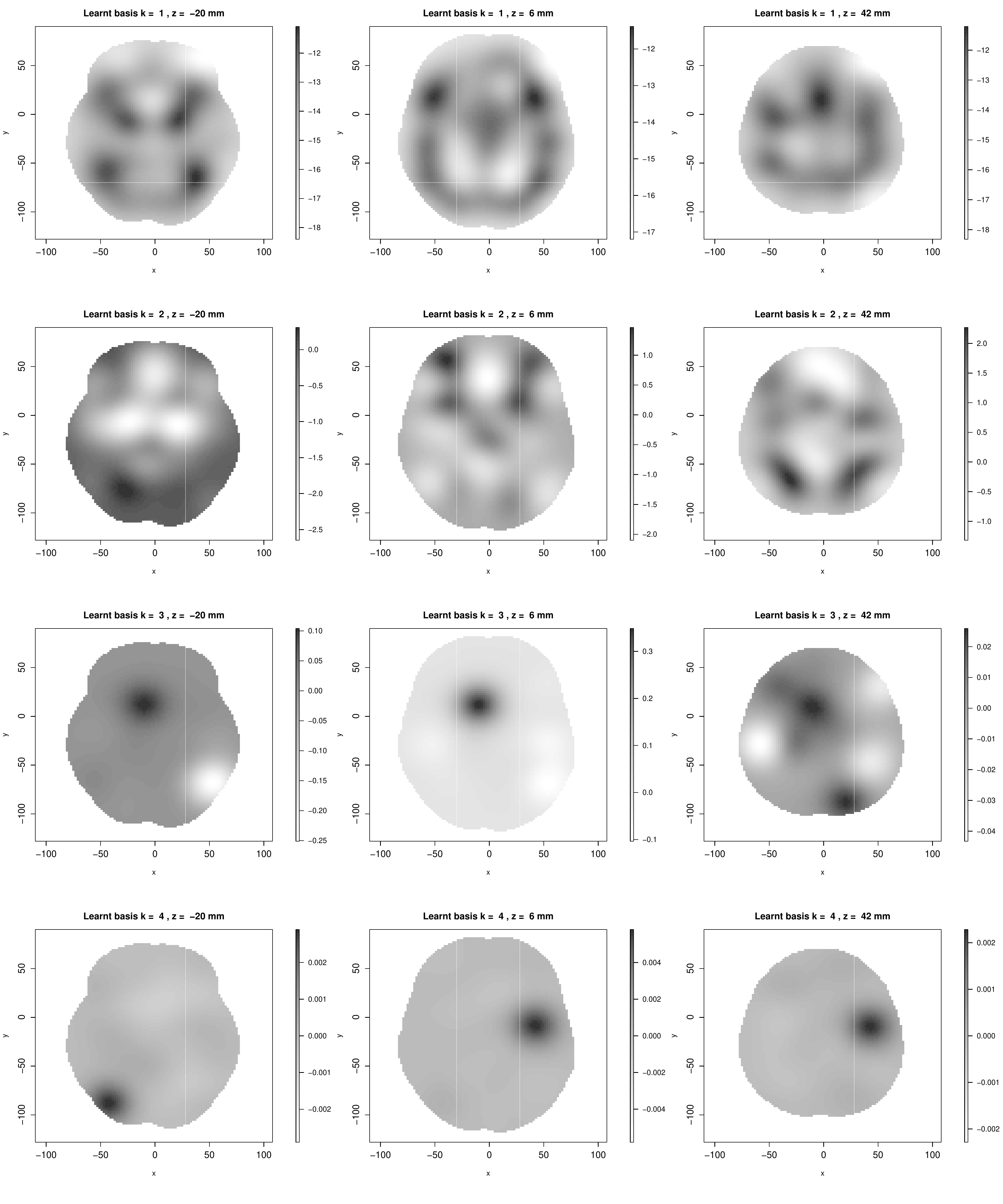}
\caption{Learnt dictionary elements $\{\phi_l\}_{l = 1}^ 4$ at three axial slices (columns). The estimated posterior mean number of factors is $k = 4$.}
\label{learntbases}
\end{figure}

\begin{figure}[h!]
\centering
\includegraphics[scale = .7]{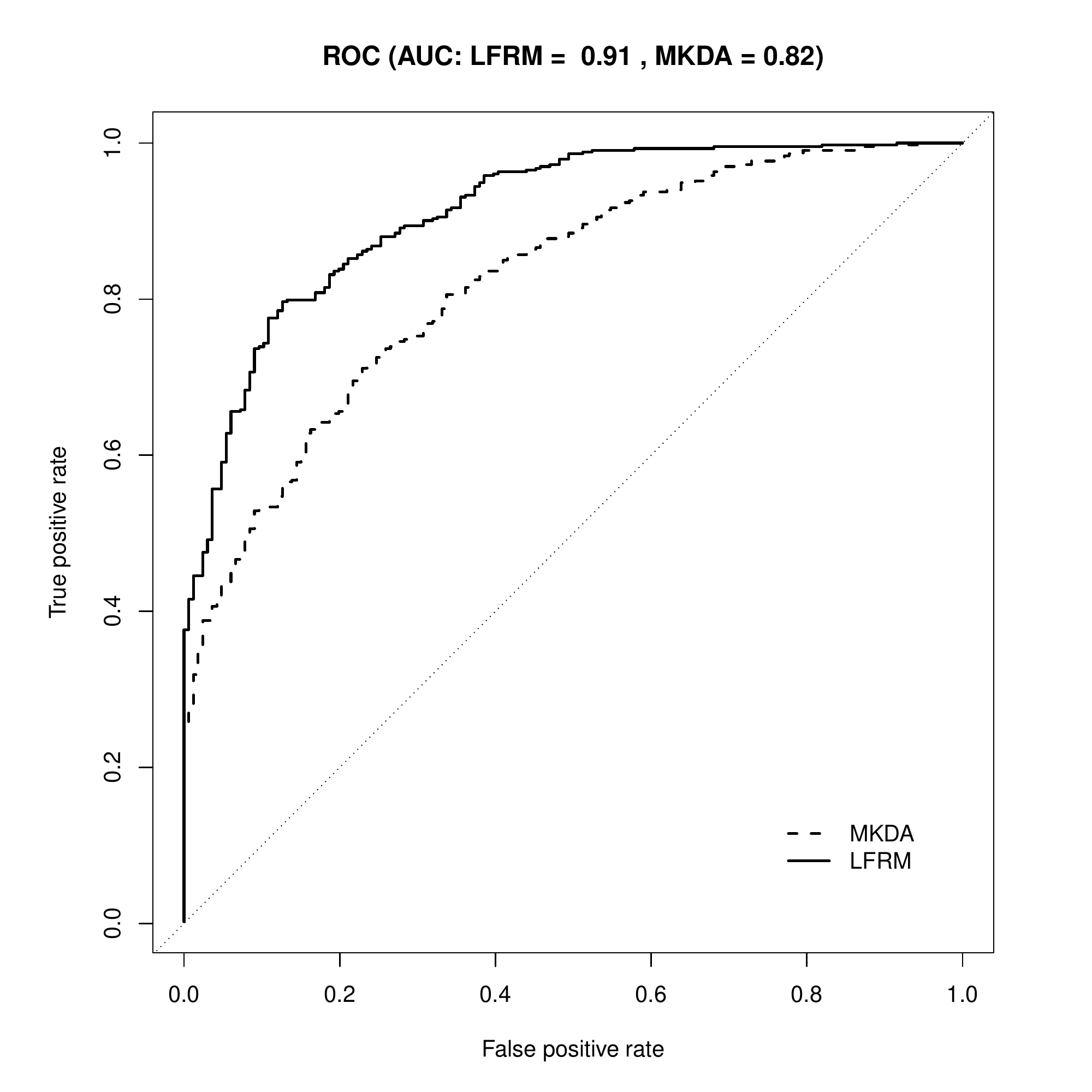}
\caption{ROC curve for prediction of study type for studies in the test set. The AUC corresponds to the area under the curve.}
\label{rocwithmua}
\end{figure}

In this Section, we apply our model to neuroimaging MA data. A typical MA dataset consists of several publications of similar, yet different experiments, and many articles report results from different statistical comparisons, here called studies. Our dataset consists of 1199 total studies categorized as either ``emotion'' (860 studies, 6481 foci; \cite{Kober2008a}) or ``executive control'' (339 studies, 4332 foci; \cite{Kim2012, Nee2007, Rottschy2012, Jonides2013}). \\ 
\indent We assigned a Gamma(1, 0.3) prior distribution with mean $1/3$ to the diagonal elements of $\bSigma^{-1}$. The hyperparametes of the MGPS prior were set to $\rho = 3$, $a_1 = 2.1$, $a_2 = 3.1$. In absence of covariate information, we assigned $\bbeta_i \sim \text{N}(\bZero, \bI)$ for $i = 1, \dots, n$. We chose $p = 352$ Gaussian kernels with bandwidth $b = 0.002$ $mm^2$. Kernels were placed on axial slices roughly 8-12 mm apart, at $z = \{-38, -22, -14,  -2,   6,  18,  28,  44\}$ mm (about 85\% of the foci were located within these axial slices) and, within each slice, were equally spaced by forming a grid of $6 \times 8$ knots along the $(x, y)$ direction. We used a standard brain mask with 2 $mm^3$ voxels and dimensions $91 \times 109 \times 91$. Kernels falling outside this mask were discarded. To update the basis function coefficients via Hamiltonian Monte Carlo \citep{Neal2011b}, we adopted the leapfrog method for $L$ steps and with a stepsize of $\epsilon$. At each iteration of the MCMC sampler, a new value for $L$ was drawn from Poisson$(25)$ and the stepsize was adapted every 10 iterations during burn-in to benchmark an average acceptance rate of 0.65 over the previous $100$ iterations in the Metropolis-Hastings step. The sampler was run for 10,000 iterations, with the first 5,000 samples discarded as a burn-in and collecting every 20$th$ sample to thin the chain. We assessed convergence of the chain by multiple runs of the algorithm from over-dispersed starting values and visually inspected the differences in the posterior mean intensity function $\mu_i(\bnu)$ at a variety of voxels and for different studies. The sampler appeared to converge rapidly and mix efficiently. \\
\indent Figure \ref{intensities} displays the posterior mean intensity function at three axial slices for two randomly chosen studies. Slices are arranged in columns, the top (bottom) row shows an emotion (executive control) study. The smoothness of the images is due to the choice of the bandwidth parameter. Larger values for $b$ (same $p$) induce rougher intensities, but results are robust and with the following conclusions unchanged. As opposed to the executive control study, the emotion study shows strong activation in two regions at slice $z = - 20$ and for $y > - 26$ mm. These regions correspond to the amygdalae; almond-shaped structures in the brain of known importance in emotion processing. The executive control study shows stronger activation in more superior slices with a bilateral pattern. Note how we identify active regions even though a study does not have foci at a particular slice or, for example, both the right and left amygdala are active for study 232 even though only one focus is reported around the right amygdala.  This a by­-product of the borrowing of information across studies in our Bayesian treatment and an essential feature to adequately estimate the intensities over the whole the brain given the sparsity of foci per study. 
\indent The top row in Figure \ref{means} shows the posterior mean difference between the estimated mean group intensity for the emotion and the executive control studies, respectively. The group intensity at iteration $t$ is obtained by averaging the basis function coefficients for studies that belong to the group, that is, 
$$ \hat{\mu}_g^t(\bnu) = \exp\{\bbb(\bnu)^\top \hat{\btheta}_g^t\}, \qquad \text{with} \qquad \hat{\btheta}_g^t = \frac{1}{\text{Card}(g)}\sum_{i \in g} \hat{\btheta}_i^t,$$
where Card$(g)$ is the cardinality of group $g$. These maps reflect the degree of consistency with which a region is activated by either task. The darker regions in the top row reveal stronger activation of emotion studies (the amygdalae at slice $z = -20$ mm), whereas the bright regions denote stronger activation of executive control studies. The grey area reveals no difference in activation. We can also obtain standard deviation maps of the difference map as a measure of uncertainty around the point estimate and the corresponding $t$-maps. The clearest of these maps is slice $z = -20$ mm where one can easily identify the amygdalae as significant regions. 
\indent It is also of interest to examine the elements of the dictionary $\{\phi_l\}_{l = 1}^k$. The interpretation of the dictionary elements has to be done with care in that they do not constitute orthogonal bases as the eigenfunctions in the fPCA literature. However, examining the dictionary  is useful to visualize how the model moves away from the fixed isotropic Gaussian kernels and learns a set of dictionary elements that are useful to represent the intensities. The posterior mean number of latent factors is $k = 4$ with 95\% credible interval $[3, 6]$. Figure \ref{learntbases} shows the first four elements of the dictionary $\{\phi_l\}_{l = 1}^4$ (columns) at axial slices $z =  -20, 6, 42$ mms (rows). Notice how the magnitude of the learnt bases decreases as $k$ increases, with the first couple of dictionary elements describing the principal patterns of activation and the later elements progressively shrunk toward zero. This effect is made possible by the MGPS prior on the factor loadings. At every axial slice, the first two dictionary elements combined recover the principal patterns of activation we observed in Figures \ref{intensities}-\ref{means}. Subsequent dictionary elements are harder to interpret and of more marginal effect. 
Our dataset is large enough to split the data into a training set (50\%), for which both foci and study type are retained for the analysis, and a testing set, for which the foci only are retained, and we test the predictive accuracy of our model. We compare our predictive method to previous work that combines MKDA and a na\"{i}ve Bayesian classifier (NBC) \citep{Yarkoni2011}. Using the MKDA framework, this method creates a study-specific binary activation map, where a voxel is given a value of 1 if it is within a 10 $mm$ (Euclidian) distance of a reported focus, and 0 otherwise. For each group (study type), an activation probability map is constructed by taking a weighted average of the binary maps of the studies in that group. Further, the predictive probability of the study type given activation from a new study is then computed using the activation probability maps via Bayes' theorem and under the assumption of independence across voxels. This method is very computationally efficient, but ignores the spatial dependence in the activation maps, leading to biased predictive probabilities of the class membership. Figure \ref{rocwithmua} shows the ROC for predicting the study type for study in the test set with corresponding area under the curve (AUC). We see that our predictive model does a better job that MKDA + NBC at predicting the emotion type. This result is robust and we could confirm it through additional chains run for sensitivity analysis to our choice for the model parameters. Therefore, taking into account the spatial information in the data helps in achieving better predictive accuracy, and our Bayesian model captures more sources of variation and conveys the uncertainty in the computation of the predictive probabilities for study type.

\section{Simulations and sensitivity analysis}
\label{s:simulation}
In this Section, we conduct simulation studies to examine how the posterior inference on the intensity functions varies with different specifications of the number of bases $p$ and bandwidth $b$, and hyperparameters $a_\sigma$, $b_\sigma$, $\alpha$, $\rho$, $a_1$, and $a_2$. In particular, we consider nine scenarios of possible combinations of $p$ and $b$ (Web Table~1) 
and, for one of this scenarios, we also report results on sensitivities to the prior specifications of the remaining parameters. 

\begin{figure}[h!]
\centering
\includegraphics[scale = .5]{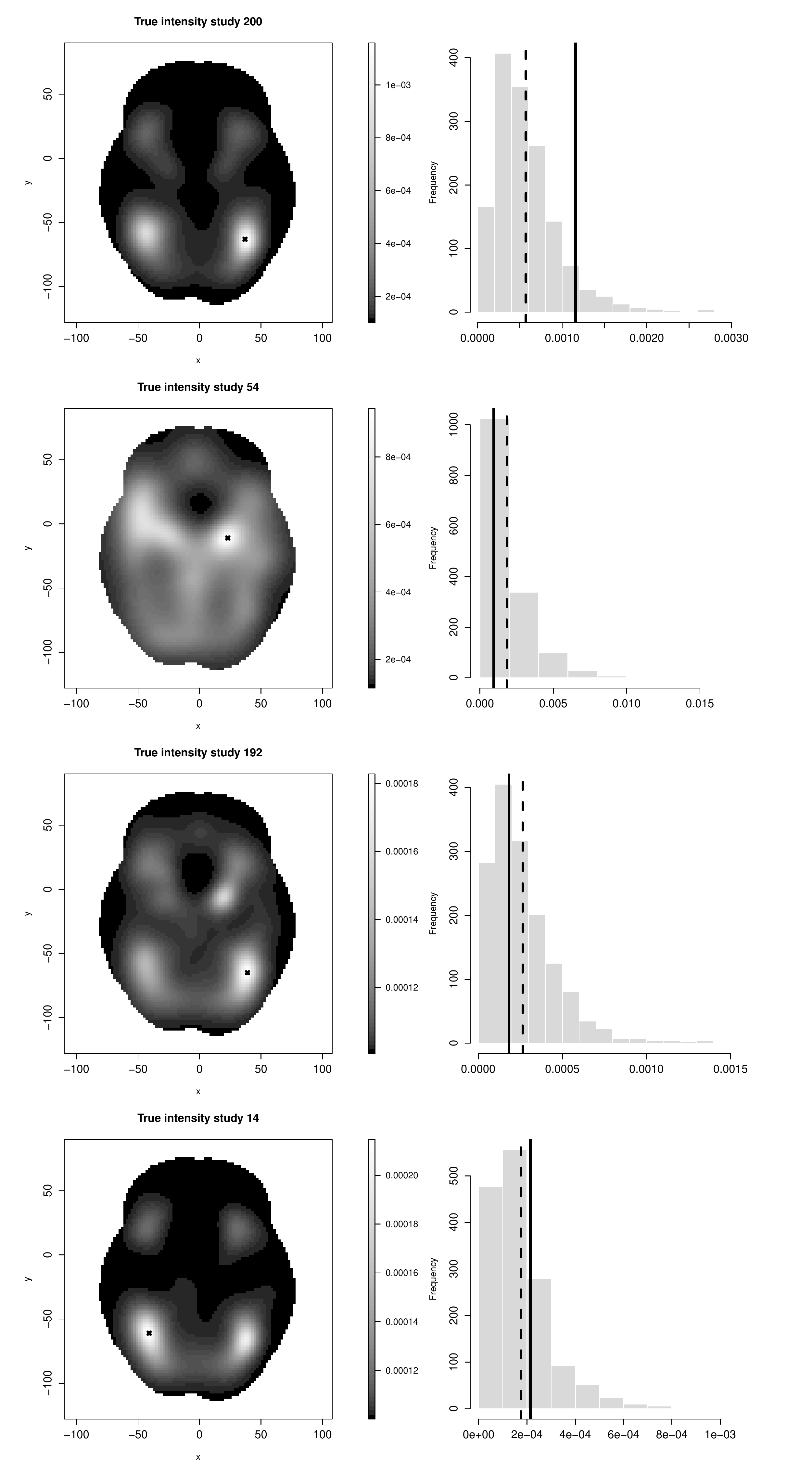}
\caption{True intensity function for four randomly chosen studies in the simulated dataset. Histograms show the posterior distribution of the intensity function evaluated at voxels of highest intensity values (black ``x''). The dashed line is the marginal posterior mean and the solid line is the true intensity. }
\label{simulation1}
\end{figure}

\indent To generate a synthetic dataset, we randomly selected 200 studies from the real data analysis (Section~\ref{neuroapp}) whilst keeping the true proportion of sampled emotion studies equal to that of the real data analysis (70\% emotion studies, hereafter called ``type 1''). We then retained the estimated posterior mean intensity functions at slice $z = - 20$ mm as true intensities for the 2D simulated dataset. Given the true intensities, the foci for each study where generated using the \textsf{spatstat} library in \textsf{R}. Web Figure~1 
shows the true group average intensity functions and the simulated data points. 
To simulate the posterior distribution, we run each chain for 50,000 iterations with a burn-in of 20,000, and thinned the chain every 20 iterations to reduce the autocorrelation in the posterior samples. To assess the posterior variability of the intensity functions, Figure \ref{simulation1} shows the posterior histograms of the intensity function evaluated at voxels with highest intensity values for four studies under case scenario 6. The true values fall in the range of posterior samples; and the posterior mean intensities are close to the true values. We examined a variety of different studies for all scenarios and conclusions were unchanged. Therefore, the method provides a good accuracy in estimating the intensities.

For a numerical comparison across the eleven scenarios, we computed the integrated mean square error (IMSE) between the posterior mean intensity function of for each study and the corresponding true function on axial slice $A$ ($z = - 20$ mm),
\begin{equation*}
\text{IMSE} = \frac{1}{N} \sum_{i = 1}^N \int_A \lambda_i(\nu) - \hat{\lambda}_i(\nu) d\nu,
\end{equation*}
where $\hat{\lambda}_i$ is the posterior mean intensity function for study $i$ and ${\lambda}_i$ is the true function. Also, we split the data into a train set (50\%) and a test set, and computed the AUC to assess the predictive accuracy. Results under the eleven scenarios are reported in Table \ref{t:z=54}. From these results, it appears that a larger number of bases with relatively narrower kernels has to be preferred to be able to capture of large variety of shapes. In general, however, the values for $p$ and $b$ have to be inferred via sensitivity analysis in that the most appropriate values for these parameters always depend on the application at hand. We observe that results under cases 9-11 are very similar. This implies that the intensity estimates are stable and not sensitive to moderate changes of the model hyperparameters. We report results of a simulation at slice $z = 42$ mm in the Supplementary Materials (Web Appendix B).

\begin{table}
\caption{Simulation study results. Comparison of the IMSE and AUC summary measures for our model under the eleven scenarios in Web Table 1. 
We report the IMSE relative to the value obtained for case scenario 9 (rIMSE).} 
\label{t:z=54}
\begin{center}
\begin{tabular}{ccc}
\hline
\hline
Scenarios & rIMSE & AUC \\ \hline
1 & 1.181 & 0.57 \\
2 & 1.186 & 0.58 \\ 
3 & 1.312 & 0.49 \\
4 & 1.147 & 0.62 \\
5 & 1.013 & 0.60 \\
6 & 1.004 & 0.65 \\
7 & 1.233 & 0.62 \\
8 & 1.122 & 0.65 \\ 
9 & \textbf{1} & \textbf{0.65} \\
10 & 1.011 & 0.65 \\
11 & 1.061 & 0.65\\
\hline
\end{tabular}
\end{center}
\end{table}

\section{Discussion}
\label{s:discuss}
The article has proposed a spatial Bayesian latent factor regression model for CBMA data. The basic formulation generalizes the Bayesian latent factor regression model of \cite{Montagna2012}, which was developed for the modeling of time-course trajectories, to the analysis of spatial point pattern data for neuroimaging MA. This allows one to include a high-dimensional set of pre-specified basis functions, while allowing automatic shrinkage and effective removal of basis coefficients not needed to characterize any of the study-specific intensity functions. Further, we accommodate joint modeling of a functional predictor with a binary response, the study type, within a framework of scalar-on-image regression. Along the same lines, the proposed framework can be easily modified for joint modeling of data of many different types, e. g., the probit model for a binary outcome can be replaced by an appropriate predictive model for categorical, nominal, or continuous study features. An interesting future direction within our modeling approach is to combine CBMA data with intensity-based meta-analysis (IBMA) data. The volume of literature on IBMA is still limited, though we note that there is a growing interest among researchers in sharing full image data and statistic maps from the studies. The extension to joint modeling of multi-type IBMA and CBMA data will be explored in future research.




\section*{Acknowledgements}

For assistance in collecting, coding, and sharing the activation coordinates used in the meta-analysis, we would like to thank Derek Nee, Simon Eickhoff, Claudia Eickhoff (Rottschy), Brian Gold, John Jonides, Ajay Satpute, Kristen Lindquist, Eliza Bliss-Moreau, and Hedy Kober, along with co-authors of the meta-analysis papers from which the coordinates were drawn. This work was partially funded by Award Number 100309/Z/12/Z from the Wellcome Trust. Dr. Johnson, Dr. Nichols, and Dr. Wager were partially supported by NIH Grant number 5-R01-NS-075066. The work presented in this manuscript represents the views of the authors and not necessarily that of the NINDS, NIH, or the Wellcome Trust Foundation.

\bibliographystyle{agsm}
 \bibliography{Functional_Neuroimaging}

\end{document}


\maketitle
\vspace{.1cm}
\begin{center}{\large{\textbf{Web Appendix A}}}\\
\vspace{.2cm}
{\textbf{Posterior computation}}\end{center}
We provide a description of the MCMC algorithm used to update from the posterior distribution of the model parameters in Section 2.1 for the LFRM on the random Poisson intensities. The sampler cycles through the following steps:

\begin{enumerate}
\item {\textit{Update of} $\btheta_i$}: Conditioning on all other model parameters, the log-posterior is given by 
	\begin{equation*}
	\log \pi(\btheta_i \mid -) 
	\propto - \int_{\mathcal{B}} \exp\{\bbb(\bs)^\top \btheta_i \}ds + \sum_{j = 1}^{n_i} \bbb(\bx_{ij})^\top\btheta_i - \frac{1}{2}(\btheta_i - \bLambda\bbeta_i)^\top \bSigma^{-1}(\btheta_i - \bLambda\bbeta_i),
	\end{equation*}
for $i = 1,\dots, n$. We recall that $\{\bx_{ij}\}_{j = 1}^{n_i}$ denotes the set of foci reported by study $i$. After discretising to a grid $B \subseteq \mathcal{B}$, the $\log$ posterior becomes
	\begin{equation} \label{logtheta} 
	\log \pi(\btheta_i \mid -) \propto - V \sum_{l \in B} \exp\{\bbb(\bl)^\top \btheta_i \} + \sum_{j = 1}^{n_i} \bbb(\bx_{ij})^\top\btheta_i  - \frac{1}{2}(\btheta_i - \bLambda\bbeta_i)^\top \bSigma^{-1}(\btheta_i - \bLambda\bbeta_i),
	\end{equation}
	where $V$ is the volume of each voxel. We resort to HMC \cite{Neal2011b} to sample from (\ref{logtheta}) by defining the potential energy function as
\[U(\btheta_i) = - \log \pi(\btheta_i \mid -),\]
and its derivative with respect to $\btheta_i$ corresponds to
	\[\frac{\partial U}{\partial \btheta_i} = V \sum_{l \in B} \bbb(\bl)^{\top} \exp\{\bbb(\bl)^{\top} \btheta_i \} -  \sum_{j = 1}^{n_i} \bbb(\bx_{ij})^{\top} + (\btheta_i - \bLambda\bbeta_i)^\top \bSigma^{-1}.\]
The kinetic energy, $K(\tilde{\btheta}_i)$, is assumed to have form $K(\tilde{\btheta}_i) = \sum_{m = 1}^p \frac{\tilde{\theta}_{ij}^2}{2}$. A detailed presentation of Hamiltonian dynamics is beyond the scope of this paper. We defer to \cite{Neal2011b} for a description of the leapfrog method and further details
\item {\textit{Update of} $\mathbf{\Lambda}$}: Sample $\lambda_{jh}, \delta_1, \delta_h, \iota_{jh}$ from the following posteriors: 
	\begin{enumerate}
				\item	Denote the $j$th row of $\mathbf{\Lambda}_{k^*}$ (the loading matrix $\mathbf{\Lambda}$ truncated to $k^* \ll p $) by $\boldsymbol{\lambda}_j$; the $\boldsymbol{\lambda}_j$'s have conditionally independent conjugate posteriors given by 
				\begin{align*}
				&\pi(\boldsymbol{\lambda}_j \mid -) \sim N_{k^*}((\mathbf{D}_j^{-1}+\sigma_j^{-2} \boldsymbol{\eta^{\top}\eta})^{-1} \boldsymbol{\eta}^{\top} \sigma_j^{-2} {\boldsymbol{\theta}^{(j)}}, (\mathbf{D}_j^{-1}+\sigma_j^{-2} \boldsymbol{\eta^{\top}\eta})^{-1})
				\end{align*}				
with $\mathbf{D}_j^{-1} = \text{diag}(\iota_{j1}\tau_1,\dots, \iota_{jk^*}\tau_{k^*})$, $\boldsymbol{\eta}^\top = [\boldsymbol{\eta}_1, \dots, \boldsymbol{\eta}_{k^*}]$ and ${\boldsymbol{\theta}^{(j)}} = (\theta_{j1}, \dots, \theta_{jn})$, for $j=1, \dots, p$ 
				\item Sample  $\iota_{jh}$  from 
				\begin{align*}
				\pi(\iota_{jh} \mid -) \sim \text{Gamma}\left(\frac{\rho + 1}{2}, \frac{\rho}{2} + \frac{\tau_h \lambda^2_{jh}}{2}\right)
				\end{align*} 
				\item Sample $\delta_1$ from
				\begin{align*}
				\pi(\delta_1 \mid -) \sim \text{Gamma}\left(a_1 + \frac{pk^*}{2}, 1 + \frac{1}{2} \sum_{l=1}^{k^*}\tau_l^{(1)}\sum_{j=1}^p \iota_{jl}\lambda_{jl}^2\right),
				\end{align*}
				and for $h \geqslant 2$, sample $\delta_h$ from 
				\begin{align*}
				\pi(\delta_h \mid -) \sim \text{Gamma}\left(a_2 + \frac{p}{2}(k^*-h+1), 1 + \frac{1}{2} \sum_{l=h}^{k^*}\tau_l^{(h)}\sum_{j=1}^p \iota_{jl}\lambda_{jl}^2\right), 
				\end{align*} 
		where $\tau_l^{(h)} = \prod_{t=1, t \neq h}^{l}\delta_t$ for $h=1,\dots,k^*$		
		\end{enumerate} 
The sampling begins with a very conservative choice of $k^*$, which is then automatically selected within the adaptive Gibbs sampler as described in \cite{Bhattacharya2011}
\item{\textit{Update of $\bbeta_i$}}: Sample $\bbeta_i$ from the full conditional posterior 
		\begin{align*}
		\pi(\bbeta_i \mid -) \sim N\left(\left(\bLambda^\top\bSigma^{-1}\bLambda + \bI_k\right)^{-1}(\bLambda^\top\bSigma^{-1}\btheta_i + \boldsymbol{\beta}^\top \bZ_i), \left(\bLambda^\top\bSigma^{-1}\bLambda + \bI_k\right)^{-1}\right)
		\end{align*}  
		for $i = 1, \dots, n$. We recall that $\bZ_i$ is the $r \times 1$ vector of covariates for study $i$
 \item {\textit{Update of} $\sigma_j^2$}: Denote with $\sigma_j^{-2},  j=1,\ldots, p,$ the diagonal elements of $\boldsymbol{\Sigma}^{-1}$. Assume $\sigma_j^{-2} \sim \text{Gamma}(a_{\sigma}, b_{\sigma})$. Sample $\sigma_j^{-2}$ from conditionally independent posteriors
			\begin{equation*}
			\pi(\sigma_j^{-2}\mid -) \sim \text{Gamma} \left(a_\sigma + \frac{n}{2}, b_\sigma + \frac{\sum_{i=1}^n({\theta}_{ij} -\boldsymbol{ \lambda_j\eta}_i)^2}{2}\right)
			\end{equation*}
		where $\blambda_j$ corresponds to the $j$th row of $\bLambda$
		\item {\textit{Update of $\boldsymbol{\beta}$:}} A Cauchy prior is induced on the columns of the $r \times k$ matrix of coefficients as follows
	\begin{equation*}
	\boldsymbol{\beta}_l \sim N(\bzr, \text{Diag}(w_{lj}^{-1})), \quad \text{with} \quad w_{lj} \sim \text{Gamma}\left(\frac{1}{2}, \frac{1}{2}\right)
	\end{equation*}
	for $l = 1, \dots, k$ and $j = 1, \dots r$. The update proceeds by sampling 
	\begin{enumerate}
	\item $\omega_{lj}$ from its full conditional posterior
		\begin{equation*}
		\pi(\omega_{lj} \mid -) \sim \text{Gamma}\left(1, \frac{1}{2}\left(1+ \beta_{lj}^2\right)\right)
		\end{equation*} 
	\item the $l$th column of $\boldsymbol{\beta}$ from its full conditional posterior
		\begin{equation*}
		\pi(\boldsymbol{\beta}_l \mid -) \sim N \left( \left(\mathbf{\tilde{Z}\tilde{Z}^{\top}}  + \mathbf{E}_l^{-1}\right)^{-1} \mathbf{\tilde{Z}} \boldsymbol{\eta}_{\cdot l}^{\top}, \left(\mathbf{\tilde{Z}\tilde{Z}^{\top}}  + \mathbf{E}_l^{-1}\right)^{-1} \right),
 		\end{equation*}
		where $\bbeta_{\cdot l}^\top \sim N(\tilde{\bZ}^{\top}\boldsymbol{\beta}_l, \bI_n)$ denotes the $l$th column of the $n\times k$ transpose of the matrix of latent factors $\bbeta$, $\bI_n$ is the $n\times n$ identity matrix, and $\tilde{\bZ}^{\top}$ denotes the transpose of the $r \times n$ matrix of predictors $\tilde{\bZ}$. Each row $i$ of $\tilde{\bZ}^{\top}$ corresponds to the vector of covariates for study $i$, for $i = 1, \dots, n$. Matrix $\mathbf{E}_l$ corresponds to $\mathbf{E}_l = \text{Diag}(\omega_{lj}^{-1})$, for $l = 1, \dots, k$ and $j = 1,\dots, r$
	\end{enumerate}
\end{enumerate}
\indent The probit extension described in Section 2.2 involves a straightforward modification of the MCMC algorithm described above, which now includes additional steps to sample from the full conditional posterior distributions of the new model parameters. To account for the study-type component model, the update of the latent factors is modified as follows 
	\begin{enumerate}
	\item[(3b)]{\textit{Update of $\bbeta_i$}}: Sample $\bbeta_i$ from a multivariate Gaussian full conditional posterior distribution with mean
	$$\mathbbm{E}(\bbeta_i \vert -) = \left(\bLambda^\top\bSigma^{-1}\bLambda + \bI_k + \bgamma\bgamma^{\top}\right)^{-1}(\bLambda^\top\bSigma^{-1}\btheta_i + \boldsymbol{\beta}^\top \bZ_i + \bgamma(W_i - \alpha)), $$
	and covariance
		$$\mathbbm{C}\text{ov}(\bbeta_i \vert -) = \left(\bLambda^\top\bSigma^{-1}\bLambda + \bI_k + \bgamma\bgamma^\top\right)^{-1}, $$
		for $i = 1, \dots, n$
		\end{enumerate}
	Further, we have the following three additional steps:
		\begin{enumerate}
	\item[(6)]{\textit{Update of $\bgamma$}}: Sample $\bgamma$ from the following full conditional posterior distribution
	$$\pi(\bgamma \vert - ) \sim N\left((\bSigma_\gamma^{-1} + \bbeta \bbeta^{\top})^{-1}(\bSigma_\gamma^{-1} \bmu_\gamma + \bbeta(\bW - \balpha)), (\bSigma_\gamma^{-1} + \bbeta \bbeta^{\top})^{-1}\right),$$
	where $\bW = [W_1, \dots, W_n]^\top$ and $\balpha$ is the $n \times 1$ vector of $n$ replicates of $\alpha$, $\balpha = [\alpha, \dots, \alpha]^\top_{n \times 1}$
	\item[(7)]{\textit{Update of $\alpha$}}: Sample $\alpha$ from the following full conditional posterior distribution
	$$\pi(\alpha \vert - ) \sim N\left(\frac{v_\alpha}{1 + v_\alpha n}\times \left(\frac{\mu_\alpha}{v_\alpha} + \sum_{i = 1}^n \left(W_i - \bbeta_i^\top\bgamma \right)\right), \frac{v_\alpha}{1 + v_\alpha n}\right)$$
	\item[(8)]{\textit{Update of $W_i$}}: Sample $W_i$ from the following full conditional posterior distribution
	\[ \pi(W_i \vert -) = 
		\begin{cases}
		TN_{[0, \infty)}(\alpha + \bgamma^\top \bbeta_i, 1) & \text{if } y_i = 1 \\
		TN_{(-\infty, 0)}(\alpha + \bgamma^\top \bbeta_i, 1) & \text{if } y_i = 0, \\
		\end{cases}
		\]
		for $i = 1, \dots, n$, where $TN_{[a, b]}$ denotes the truncated normal distribution on the interval $[a, b]$
	\end{enumerate}

\vspace{1cm}
\begin{center}{\large{\textbf{Web Appendix B}}}\\
\vspace{.2cm}
{\textbf{Simulation study (cont.`d Section 4)}}\end{center}

\begin{figure}[h!]
\centering
\includegraphics[scale = .55]{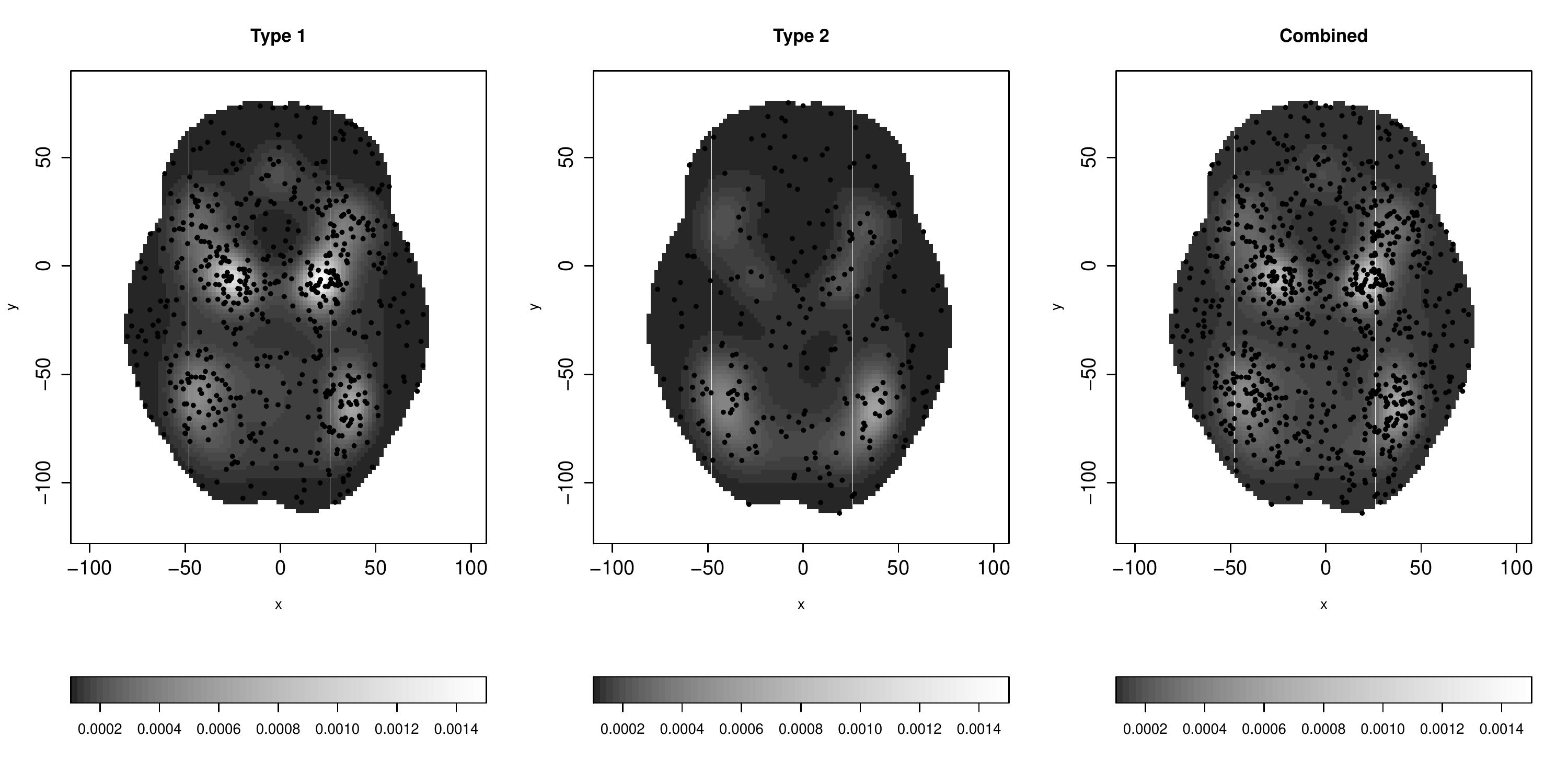}
\caption{True group average intensity functions with data points for the simulation study in Section 4 of the paper.}
\label{synthdata}
\end{figure}

\noindent Web Table~\ref{t:one} and Web Figure~\ref{synthdata} refer to the simulation study presented in Section 4 of the paper. \\

\begin{table}[h!]
\caption{Prior specifications of eleven scenarios for sensitivity analysis.}
\label{t:one}
\begin{center}
\begin{tabular}{ccccccccc}
\hline
\hline
Scenarios & $p$ & $b$ & $a_\sigma$ & $b_\sigma$ & $m_\alpha$ & $a_1$ & $a_2$ & $\rho$ \\ \hline
1 & 25 & 1/800 & 1 & 0.3 & $\Phi^{-1}(0.5)$ & 2.1 & 3.1 & 3 \\
2 & 25 & 1/512 & \multicolumn{6}{c}{$\vdots$} \\ 
3 & 25 & 1/128 & \multicolumn{6}{c}{$\vdots$} \\
4 & 48 & 1/800 & \multicolumn{6}{c}{$\vdots$} \\
5 & 48 & 1/512 & \multicolumn{6}{c}{$\vdots$} \\
6 & 48 & 1/128 & \multicolumn{6}{c}{$\vdots$} \\
7 & 90 & 1/800 & \multicolumn{6}{c}{$\vdots$} \\
8 & 90 & 1/512 & \multicolumn{6}{c}{$\vdots$} \\ 
9 & 90 & 1/128 & 1 & 0.3 & $\Phi^{-1}(0.5)$ & 2.1 & 3.1 & 3 \\
10 & 90 & 1/128 & 1 & 0.25 & $\Phi^{-1}(0.62)$ & 3.1 & 2.1 & 2 \\
11 & 90 & 1/128 & 2 & 2 & $\Phi^{-1}(0.62)$ & 3.1 & 2.1 & 2 \\
\hline
\end{tabular}
\end{center}
\end{table}

\vspace{.2cm}
\begin{center}{\textbf{Sensitivity analysis}}\end{center}

\begin{figure}[h!]
\centering
\includegraphics[scale = .55]{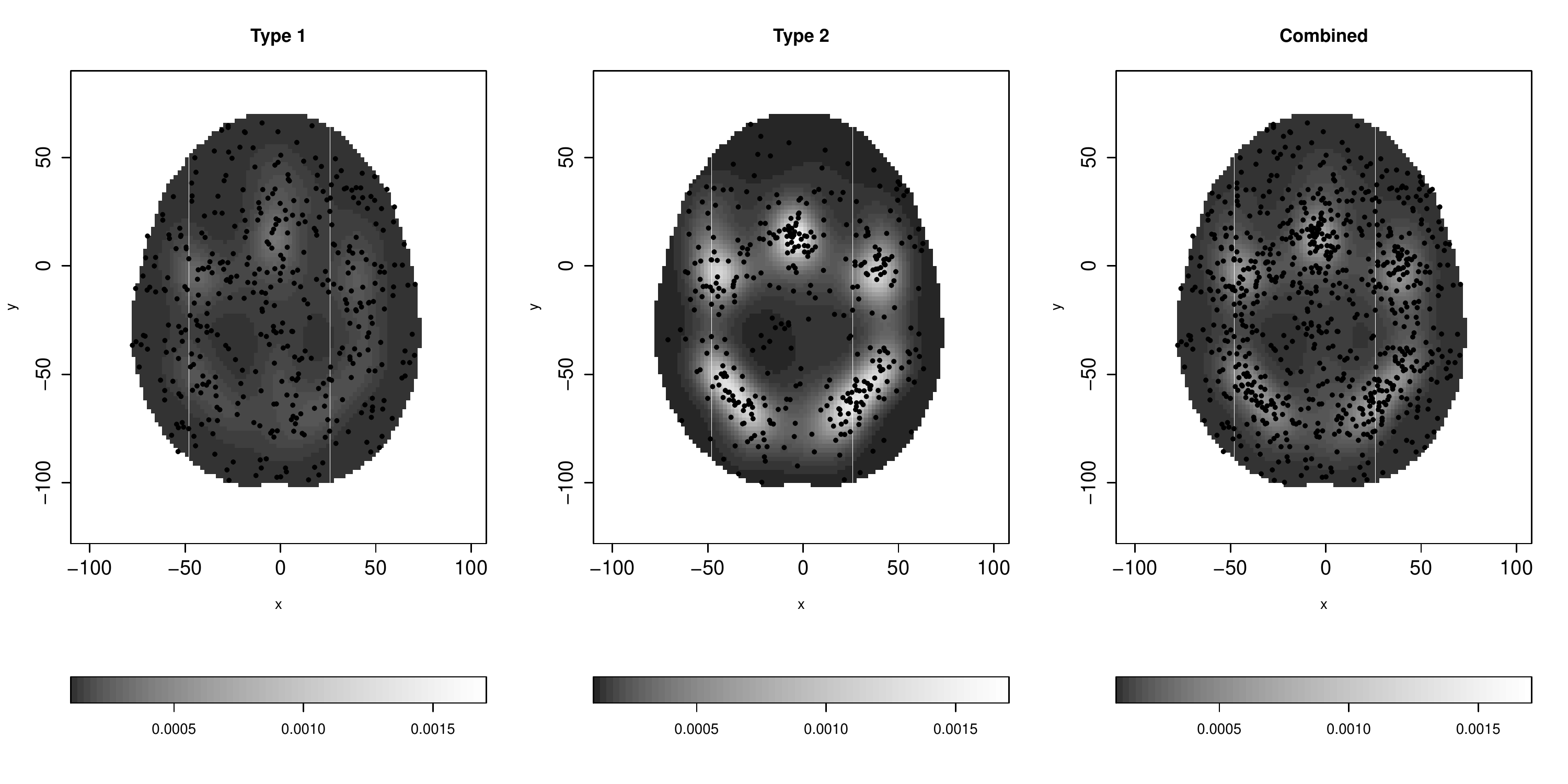}
\caption{True group average intensity functions with simulated data points at slice $z = 42$ mm.}
\label{synthdata2}
\end{figure}

\noindent We report results on a second simulation study for sensitivity analysis. Data was generated as described in Section~4, but we now retained the estimated marginal posterior mean intensity functions at slice $z = 42$ mm as true intensities for the 2D dataset. Web Figure \ref{synthdata2} shows the true group average intensity functions and simulated data points. 

\begin{figure}[h!]
\centering
\includegraphics[scale = .45]{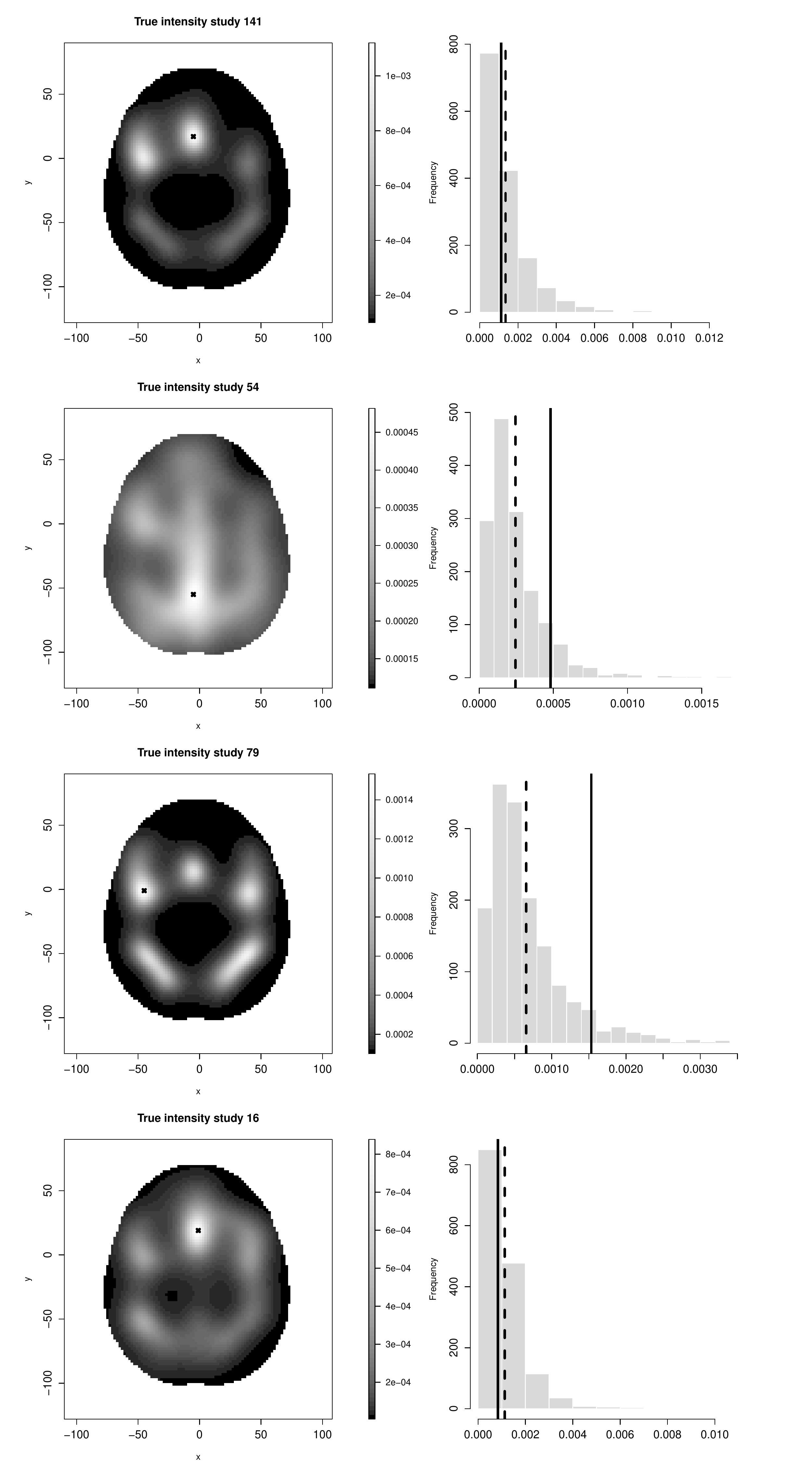}
\caption{True intensity function for four randomly chosen studies in the simulated dataset at $z = 42$ mm. Histograms show the posterior distribution of the intensity function evaluated at voxels of highest intensity values (black ``x''). The dashed line is the marginal posterior mean and the solid line is the true intensity. }
\label{simulation2}
\end{figure}

\indent To simulate the posterior distribution, we run the chain for 50,000 iterations with a burn-in of 20,000, and thinned the chain every 20 iterations to reduce the autocorrelation in the posterior samples. To assess the posterior variability of the intensity functions, Web Figure \ref{simulation2} shows the posterior histograms of the intensity function evaluated at voxels with highest intensity values for four studies. The true values fall in the range of posterior samples; and the posterior mean intensities are close to the true values. We examined a variety of other different studies and conclusions were unchanged. Therefore, the method provides a good accuracy in estimating the intensities.

Web Table~\ref{t:z=116} shows a comparison of the relative IMSE and AUC summary measures for our model under nine scenarios of possible combinations of $p$ and $b$. Again, the studies were split into a train set (50\%) and a test set to assess the predictive accuracy. As we observed in Section~4, large values of $p$ with relatively narrow kernels are to be preferred to obtain a reduction in IMSE to the truth. However, the AUC does not vary significantly across the nine scenarios, and it is higher than the AUC obtained for the simulations at slice $z = - 20$ mm. A possible is explanation is that the foci are more evenly spread between the two study types at this slice (Web Figure~\ref{synthdata2}), whereas the type 1 studies in Section~4 have more foci as compared to type 2 (Web Figure~\ref{synthdata}), and therefore the predictive classification is more likely to be biased toward type 1.

\begin{table}
\caption{Simulation study results ($z = 42$ mm). Comparison of the IMSE and AUC summary measures for our model under nine scenarios of possible combinations of $p$ and $b$. We report the IMSE relative to the value obtained for case scenario 9 (rIMSE).} 
\label{t:z=116}
\begin{center}
\begin{tabular}{ccccc}
\hline
\hline
Scenarios & $p$ & $b$ & rIMSE & AUC \\ \hline
1 & 21 & 1/800 & 1.230 & 0.74\\
2 & 21 & 1/512 & 1.272 & 0.72\\ 
3 & 21 & 1/128 & 1.413 & 0.75\\
4 & 45 & 1/800 & 1.115 & 0.73\\
5 & 45 & 1/512 & 1.034 & 0.74\\
6 & 45 & 1/128 & 1.157 & 0.73\\
7 & 86 & 1/800 & 1.264 & 0.72\\
8 & 86 & 1/512 & 1.146 & 0.74 \\ 
9 & 86 & 1/128 & \textbf{1} & \textbf{0.74}\\
\hline
\end{tabular}
\end{center}
\end{table}

\bibliographystyle{plain}
\bibliography{Functional_Neuroimaging}